\documentclass[aps,english,prd,twocolumn,superscriptaddress,nofootinbib,preprintnumbers,10pt,notitlepage,eqsecnum]{revtex4}
\usepackage[latin1]{inputenc}
\usepackage{graphicx}
\usepackage{amsfonts}
\usepackage{amsmath,amsthm,amssymb}
\usepackage{enumitem}
\usepackage{bm}
\usepackage{graphicx}
\usepackage{amsfonts}
\usepackage{dcolumn}
\usepackage{color}
\usepackage{mathrsfs}

\def\be{\begin{equation}}
\def\ee{\end{equation}}
\def\bea{\begin{eqnarray}}
\def\eea{\end{eqnarray}}
\def\bs{\begin{subequations}}
\def\es{\end{subequations}}

\def\a{\alpha}
\def\b{\beta}

\def\vf{\varphi}

\def\m{\mu}
\def\n{\nu}

\def\o{\omega}
\def\O{\Omega}

\def\nn{\nonumber}

\def\cL{\mathcal{L}}

\usepackage{color}

\makeatother

\usepackage{babel}
\makeatother

\begin{document}

\title{Planck constraints on inflation in auxiliary vector modified $f(R)$ theories}

\author{Mehmet Ozkan}
\affiliation{Van Swinderen Institute for Particle Physics and Gravity,
University of Groningen, Nijenborgh 4, 9747 AG Groningen, The Netherlands}

\author{Yi Pang}
\affiliation{George and Cynthia Woods Mitchell Institute for Fundamental Physics and Astronomy,
Texas A\&M University, College Station, TX 77843, USA}

\author{Shinji Tsujikawa}
\affiliation{Department of Physics, Faculty of Science, Tokyo University of Science, 1-3,
Kagurazaka, Shinjuku, Tokyo 162-8601, Japan}

\date{\today}

\begin{abstract}

We show that the universal $\alpha$-attractor models of inflation can be realized
by including an auxiliary vector field $A_{\mu}$ for the Starobinsky
model with the Lagrangian $f(R)=R+R^2/(6M^2)$.
If the same procedure is applied to general modified $f(R)$
theories in which the Ricci scalar $R$ is replaced by
$R+A_{\mu} A^{\mu}+\beta \nabla_{\mu}A^{\mu}$ with constant $\beta$,
we obtain the Brans-Dicke theory with a scalar
potential and the Brans-Dicke parameter $\omega_{\rm BD}=\beta^2/4$.
We also place observational constraints on inflationary models
based on auxiliary vector modified $f(R)$ theories from the latest
Planck measurements of the Cosmic Microwave Background (CMB)
anisotropies in both temperature and polarization.
In the modified Starobinsky model, we find that the parameter $\beta$
is constrained to be $\beta<25$ (68\,\%\,confidence level) from the bounds
of the scalar spectral index and the tensor-to-scalar ratio.

\end{abstract}

\maketitle


\section{Introduction}
\label{intro}

The inflationary paradigm \cite{Stamodel,oldinf} has been the backbone of
high energy cosmology over the past three decades.
For the realization of inflation, we require the existence of
at least one additional degree of freedom to the Einstein-Hilbert action.
A canonical scalar field with a nearly flat potential
can play such a role \cite{newinf,chaotic}.
In modified gravitational theories, a scalar
degree of freedom generally emerges as a result of the breaking
of gauge symmetries present in
General Relativity \cite{Faraoni,fRreview,Tsujilec,Clif}.

The first model of inflation, which was proposed by Starobinsky
in 1979 \cite{Stamodel}, is based on the modification of gravity with the Lagrangian
$f(R)=R+R^2/(6M^2)$, where $R$ is the Ricci scalar and $M$ is
a constant having a dimension of mass.
In general, the $f(R)$ gravity is equivalent to
the Brans-Dicke (BD) theory \cite{Brans} with the BD parameter
$\omega_{\rm BD}=0$ \cite{Ohanlon}.
The propagation of a scalar degree of freedom in $f(R)$ gravity is
particularly transparent in the Einstein frame where a canonical scalar field $\phi$
evolves along a potential of gravitational origin \cite{conformal}.

The Starobinsky model gives rise to the Einstein-frame
potential with a nearly flat region responsible
for inflation \cite{fRreview,Maeda}.
In this model the scalar spectral index $n_s$ and the tensor-to-scalar
ratio $r$ of primordial density perturbations generated during inflation
are given, respectively, by $n_s \simeq 1-2/N$ and
$r \simeq 12/N^2$, where $N$ is the number of e-foldings
on scales relevant to the CMB
temperature anisotropies \cite{Kofman,Feldman,Hwang}.
The Starobinsky model is consistent with the recent joint analysis
of the Planck temperature data \cite{Planckcosmo} and
the B-mode polarization data from BICEP2/Keck array with
the Planck maps at higher frequencies \cite{BKP}.
The tensor-to-scalar ratio is constrained to be $r<0.08$ at
95\,\% confidence level (CL) from such a joint analysis.
The large-field models like chaotic inflation and natural
inflation are now in tension with the CMB data \cite{Planckinf}.

Recently, there have been numerous attempts to embed
the Starobinsky model in the framework of
supergravity \cite{Ketov}-\cite{Ozkan}
or ghost-free higher-derivative gravitational
theories \cite{Biswas,Modesto}.
The bottom line is how to build up the Einstein-frame potential
similar to the form $V(\phi)=V_0 (1-e^{-\sqrt{2/3}\,\phi/M_{\rm pl}})^2$,
where $M_{\rm pl}=2.435 \times 10^{18}$~GeV is the reduced Planck mass.
In a particular version of supergravity where the inflaton is a part
of a vector multiplet it is possible to construct a generalized
potential of the form $V(\phi)=V_0 (1-e^{-\sqrt{2/(3\alpha)}\,\phi/M_{\rm pl}})^2$,
where the parameter $\alpha$ is inversely proportional to
the curvature of the inflaton K\"ahler manifold \cite{KLR}.
This was dubbed the $\alpha$-attractor model in which
the inflationary period is followed by the reheating stage
with the inflaton oscillations around $\phi=0$.

In the limit $\alpha \to \infty$ the potential of the
$\alpha$-attractor model is approximately given by
$V(\phi) \propto \phi^2$, so it is equivalent to that of
the quadratic potential in chaotic inflation \cite{chaotic}.
For $1 \le \alpha<\infty$ the tensor-to-scalar
ratio is in the range $O(10^{-3})<r<O(10^{-1})$,
with $n_s$ inside the 95\,\%\,CL observational
contour constrained by the Planck
data \cite{Planck1,Kuro,Planckinf}.
In Ref.~\cite{Tavakol} the authors derived the
same potential as that of the $\alpha$-attractor model
by generalizing the Starobinsky model in the
framework of the BD theory and they placed observational
constraints on the model from the WMAP 7yr data.

In this paper we show that the $\alpha$-attractor model
arises by introducing an auxiliary vector field $A_{\mu}$
and replacing the Ricci scalar $R$ with
$R+A_{\mu}A^{\mu}+\beta \nabla_{\mu}A^{\mu}$
in the Starobinsky Lagrangian, where $\nabla_{\mu}$
is the covariant derivative.
This procedure can be extended to the general $f(R)$ Lagrangian.
The resulting theory is equivalent to the BD theory characterized
by the BD parameter $\omega_{\rm BD}=\beta^2/4$
with one scalar propagating degree of freedom.

In light of the recent release of the Planck temperature
and polarization data, we also put observational constraints
on inflationary models in the framework of auxiliary vector
modified $f(R)$ theories. Our analysis not only encompasses
the $\alpha$-attractor model but also the models derived by
promoting the Lagrangian $f(R)=R+cR^n$ ($n>1$) to
include the auxiliary vector field.
This can accommodate a wider class of inflationary models
including chaotic inflation with the
potential $V(\phi) \propto \phi^{n/(n-1)}$.

This paper is organized as follows.
In Sec.~\ref{stamodel} we review the Starobinsky model
and its dual description in terms of a scalar degree
of freedom $\phi$.
In Sec.~\ref{auximodel} we show how the $\alpha$-attractor
model emerges by modifying the Starobinsky model with
inclusion of the auxiliary vector field.
In Sec.~\ref{mofR} we extend this prescription
to general $f(R)$ theories and provide the formulas
of inflationary observables associated with the primordial
scalar and tensor perturbations.
In Sec.~\ref{obcon} we place observational bounds on the
auxiliary vector modified inflationary models from
the latest Planck data combined with other
B-mode polarization data.
Sec.\,\ref{consec} is devoted to conclusions.

\section{Starobinsky model and its dual description}
\label{stamodel}

The Starobinsky model \cite{Stamodel} is described by the action
\be
S=\frac{M_{\rm pl}^2}{2} \int d^4 x \sqrt{-g}\,{\cal L}(R)\,,
\label{Str}
\ee
where $g$ is the determinant of the space-time
metric $g_{\mu \nu}$ and ${\cal L}(R)$ is
a function of $R$ of the form
\be
{\cal L}(R)= R + \frac{R^2}{6M^2}\,.
\label{LRSta}
\ee

The discussion given below is already well known 
in the literature \cite{fRreview}, but this is useful for the comparison with 
the auxiliary modified Starobinsky model given in 
Sec.~\ref{auximodel}.  
The model (\ref{LRSta}) possesses an additional scalar degree
of freedom to that in General Relativity.
In order to make this manifest,
we consider the following Lagrangian
\be
\cL = F  + \frac{F^2}{6M^2}- \vf (F - R)\,.
\label{dual1}
\ee
It is easy to see that, upon integrating out the field $\vf$,
we get back to the original Starobinsky model (\ref{Str}).
Varying Eq.~(\ref{dual1})
with respect to $F$, it follows that
\be
F = 3 M^2 (\vf-1)\,.
\ee
Then, the Lagrangian (\ref{dual1}) can be rewritten as
\be
\cL = \vf R - \frac32 M^2 (\vf-1)^2 \,.
\label{fRBD}
\ee
This is equivalent to the BD theory \cite{Brans}
with the BD parameter $\omega_{\rm BD}=0$ and
the scalar potential $V(\vf)=(3/2)M^2 (\vf-1)^2$.

In Eq.~(\ref{fRBD}) the scalar degree of freedom $\vf$ is directly
coupled to the Ricci scalar $R$.
One can transform the action (\ref{Str}) with the Lagrangian (\ref{fRBD})
to the so-called Einstein frame under the conformal transformation
$\widetilde{g}_{\m\n} = \O^2 (\varphi) g_{\m\n}$ \cite{conformal}.
Denoting the quantities in the transformed frame as a tilde,
we have the following relations \cite{fRreview}
\bea
\sqrt{-g} &=& \O^{-4} \sqrt{-\widetilde{g}}\,,\\
R &=& \O^2 (\widetilde{R} + 6 \widetilde{\Box}\o
-6 \widetilde{g}^{\m\n} \nabla_\m \o \nabla_\n \o)\,,
\label{Rtra}
\eea
where $\o \equiv \ln \O$.
We obtain the Einstein-frame action for the choice
\be
\Omega^2=\vf\,,
\label{Omechoice}
\ee
under which the Ricci scalar $\widetilde{R}$ does not
have a direct coupling with $\vf$.
Dropping the total derivative term $\widetilde{\Box}\o$
in Eq.~(\ref{Rtra}) and introducing a scalar field
\be
\phi \equiv \sqrt{\frac32}M_{\rm pl}\,\ln \vf\,,
\ee
the action in the Einstein frame reads
\be
S=\int d^4 x \sqrt{-\widetilde{g}} \left[
\frac{M_{\rm pl}^2}{2} \widetilde{R}-\frac12
\widetilde{g}^{\mu \nu} \nabla_{\mu} \phi
\nabla_{\nu} \phi-V(\phi) \right]\,,
\label{actionEin}
\ee
where
\be
V(\phi)=\frac34 M_{\rm pl}^2 M^2
\left( 1-e^{-\sqrt{\frac23} \frac{\phi}{M_{\rm pl}}}
\right)^2\,.
\label{Stapotential}
\ee
Hence the scalar degree of freedom $\phi$, which
is the gravitational origin, propagates with
the kinetic energy
$-(1/2)\widetilde{g}^{\mu \nu} \nabla_{\mu} \phi
\nabla_{\nu}\phi$.
The potential (\ref{Stapotential}) is sufficiently flat for
$\phi$ larger than the order of $M_{\rm pl}$,
in which regime inflation
occurs due to the slow-roll evolution of $\phi$.

\section{Auxiliary Modified Starobinsky Model}
\label{auximodel}

The auxiliary vector modified Starobinsky model is inspired 
by the supersymmetric extension of 
Starobinsky model \cite{Ketov}-\cite{Ozkan}. 
In the old minimal formulation of ${\cal N}=1$ off-shell supergravity, 
the Weyl multiplet consists of the vielbein $e^a_{\mu}$, 
the gravitino $\psi^a_{\mu}$, an auxiliary vector $A_{\mu}$ 
and an auxiliary complex scalar $S$. 
The embedding of Starobinsky model in the old minimal supergravity 
is obtained by coupling a chiral multiplet to the Weyl multiplet. 
The supersymmetric Starobinsky model can be recast into the form of 
a scalar-tensor theory by integrating out the auxiliary fields. 
In particular, integrating out the auxiliary vector field generates 
the kinetic term for the imaginary part of the complex scalar 
in the chiral multiplet\footnote{A detailed performance of this procedure 
can be found in the Sec.~5 of \cite{Ozkan}.}. 

We apply the similar mechanism here by coupling an auxiliary vector 
field to the Starobinsky model in a specific way such that the auxiliary 
vector field does not generate new degrees of freedom. 
As we will show below, integrating out the auxiliary vector field modifies 
the kinetic term of inflaton. 
The resulting theory written in the Einstein frame coincides with 
the $\alpha$-attractor model proposed in Ref.~\cite{KLR}. 
In analogous to the supersymmetric extension of Starobinsky model, 
the auxiliary vector modified Starobinsky model thus provides 
a gravitational origin for the designed scalar potential in 
the $\alpha$-attractor model. 
We can also apply the same mechanism to auxiliary vector 
coupled $f(R)$ theories and obtain a class of 
generalized $\alpha$-attractor models (see Sec.~\ref{auxfRsec}).

We would like to stress that our model is inspired by the ${\cal N}=1$ 
off-shell supergravity, but it does not directly come from a supersymmetric 
scenario with a SUSY breaking mechanism. Hence we do not take into account 
the effect of gravitinos for the cosmological dynamics.
Construction of a SUSY breaking $\alpha$-attractor model with the effect 
of gravitinos taken into account is beyond the scope of our paper.

We start with the Lagrangian of the form
${\cal L}=R+ A_\m A^\m + \b \nabla_\m A^\m$, where
$\beta$ is a constant. Note that the coefficient in front of the term
$A_\m A^\m$ has been fixed to 1.
When higher derivative terms are included, $A_\m$ can pick up
kinetic terms such as $(\nabla_\m A^\m)^2$ and $F_{\m\n} F^{\m\n}$,
so the auxiliary vector starts to propagate. If we would like to allow
kinetic terms for $A_\m$ but still wish to keep $A_\m$
as an auxiliary vector in the higher derivative extended model, then
the action of the auxiliary vector modified Starobinsky model has
to take the following form:
\be
S=\frac{M_{\rm pl}^2}{2}\int d^4 x \sqrt{-g}\,{\cal L}\,,
\label{Sauxi}
\ee
where
\bea
\cL &=& R +  A_\m A^\m + \b \nabla_\m A^\m \nn\\
&& + \frac{1}{6M^2} \Big( R +  A_\m A^\m + \b \nabla_\m A^\m \Big)^2 \,.
\label{aS1}
\eea

In order to see that this model gives rise to only one scalar degree
of freedom, we perform the similar analysis to that performed
in the previous section.
We first write Eq.~(\ref{aS1}) as
\be
\cL = F + \frac{1}{6M^2} F^2 -
\vf \Big( F- R - A_\m A^\m - \b \nabla_\m A^\m \Big) \,.
\label{lag}
\ee
Varying the Lagrangian (\ref{lag}) with respect to $A^{\mu}$
and $F$, respectively, it follows that
\bea
A_\m &=& \frac1{2\vf} \b \nabla_\m \vf \,,\label{Am} \\
F &=& 3 M^2 (\vf-1)\,.\label{Fm}
\eea
The equation of motion (\ref{Am}) of the auxiliary vector field 
demonstrates that on-shell, the vector field is equivalent to the 
gradient of the scalar field. Therefore, 
there are no dynamical spin-1 degrees of freedom in our model. 
Consequently, the apparent presence of
a vector field in our model does not spoil the homogeneity
and isotropy of the universe.

Substituting the relations (\ref{Am})-(\ref{Fm}) into Eq.~(\ref{lag}) and dropping
a total derivative term, we obtain the (Jordan-frame) Lagrangian
\be
\cL = \vf R - \frac{1}{4\vf} \b^2 \nabla_\m \vf \nabla^\m \vf
- \frac32 M^2 (\vf-1)^2 \,.
\label{Lpo}
\ee
This theory is equivalent to the BD theory
with the BD parameter
\be
\omega_{\rm BD}=\frac14 \beta^2\,,
\label{omega}
\ee
so the auxiliary vector model
(\ref{aS1}) possesses one scalar degree
of freedom.

Under the conformal transformation
$\widetilde{g}_{\m\n} = \O^2(\vf) g_{\m\n}$
with $\Omega^2=\vf$,
the action in the Einstein frame reads
\be
S=\int d^4 x \sqrt{-\widetilde{g}} \left[
\frac{M_{\rm pl}^2}{2} \widetilde{R}-\frac12
\widetilde{g}^{\mu \nu} \nabla_{\mu} \phi
\nabla_{\nu} \phi-V(\phi) \right]\,,
\label{Ein}
\ee
where $\phi$ is a canonical scalar field defined by
\be
\phi=\frac{\sqrt{6+\beta^2}}{2}M_{\rm pl} \ln \vf\,.
\label{phidef}
\ee
The potential $V(\phi)$ is given by
\be
V(\phi)=\frac34 M_{\rm pl}^2 M^2
\left( 1-e^{-\sqrt{\frac{2}{3\alpha}}
\frac{\phi}{M_{\rm pl}}} \right)^2\,,
\label{alpo}
\ee
where
\be
\alpha \equiv 1+\frac{\beta^2}{6}
=1+\frac23 \omega_{\rm BD}\,.
\label{aldef}
\ee
This is equivalent to the $\alpha$-attractor model studied
in Ref.~\cite{KLR}. Setting $\beta=0$, we recover the Starobinsky 
model described by the potential (\ref{Stapotential}) in the Einstein frame.

\begin{figure}
\centering \noindent
\includegraphics[width=3.5in,height=3.5in]{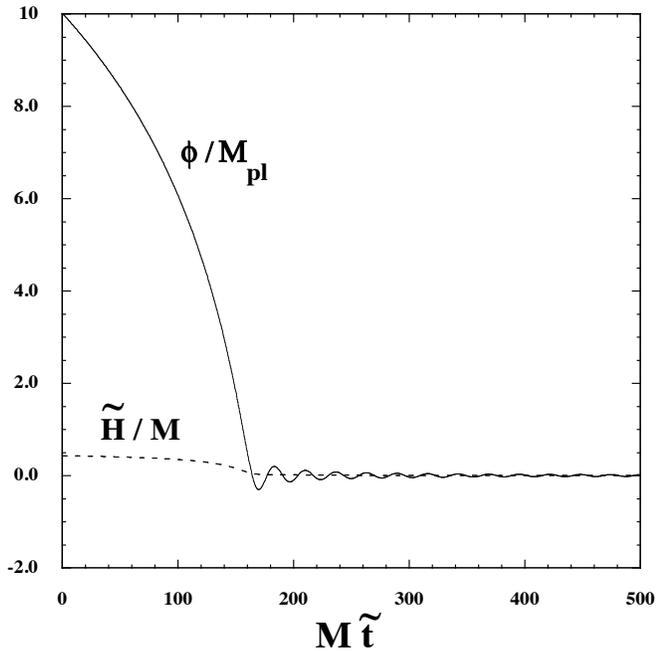}
\caption{Evolution of the scalar field $\phi$ and the Hubble 
parameter $\tilde{H}$ in the Einstein frame for $\beta=10$. 
We choose the initial conditions $\phi=10M_{\rm pl}$ and 
$\dot{\phi}=0$ at $\tilde{t}=0$. In this case the inflationary 
period lasts with the number of e-foldings about 55, 
which is followed by the reheating stage with the 
oscillating inflaton field.}
\label{fig1}
\end{figure}

Around $\phi=0$, the potential (\ref{alpo}) is approximately given by 
$V(\phi) \simeq M^2 \phi^2/(2\alpha)$.
The graceful exit from inflation to reheating naturally occurs 
after the field $\phi$ enters the regime 
$\sqrt{2/(3\alpha)}\phi/M_{\rm pl} \ll 1$. 
In Fig.~\ref{fig1} we plot the evolution of the field $\phi$ and the 
Hubble parameter $\tilde{H}$ versus the cosmic time $\tilde{t}$ 
in the Einstein frame for $\beta=10$. 
During inflation, the field evolves slowly along the potential (\ref{alpo}) 
with a nearly constant Hubble parameter.  
For the parameter $\beta$ and initial conditions chosen in Fig.~\ref{fig1},  
the end of inflation is characterized by the field value 
$\phi_f \simeq 0.8M_{\rm pl}$ with the number of e-foldings 
$N \simeq 55$. This shows good agreement with the analytic estimation 
given in Sec.~\ref{ausec} [see Eq.~(\ref{Nes})]. 

As we see in Fig.~\ref{fig1}, the Universe exits from the
inflationary epoch to the reheating stage driven by the oscillation 
of $\phi$. The field $\phi$ exhibits a damped oscillation around
the potential minimum at $\phi=0$. 
Since the auxiliary vector only gives rise to the change of 
the kinetic term of $\varphi$ in Eq.~(\ref{Lpo}), 
it does not modify the cosmological dynamics after the field 
$\phi$ stabilizes at the minimum of the Einstein-frame 
potential (\ref{alpo}), i.e., after reheating. 
In Appendix A we estimate the time at the onset of radiation-dominated 
era. After this epoch, the energy density of radiation dominates over
that of $\phi$.

Since the field $\phi$ behaves as a massive oscillating scalar around 
the potential minimum, the basic mechanism of reheating 
is similar to that in the Starobinsky model \cite{reheating,reheating2,fRreview} 
apart from the fact that the energy scale of the potential gets 
lowered by the factor $1/\alpha$. 
The modified kinetic term in the Jordan frame can 
be interpreted as the modified shape of the potential in 
the Einstein frame. As we will see in Sec.~\ref{ausec}, 
this modification of the Einstein-frame potential gives rise to the change of  
CMB observables relative to those in the Starobinsky model.
Especially, the larger value of the tensor-to-scalar ratio $r$ 
caused by the modification of the kinetic term of $\varphi$ is 
a distinguished observational feature to discriminate between 
the $\alpha$-attractor model [$O(10^{-3})<r<O(10^{-1})$] 
and the Starobinsky model [$r=O(10^{-3})$].

The observational constraints on the potential (\ref{alpo})
were discussed in Ref.~\cite{Tavakol} with the
WMAP 7yr data and in Ref.~\cite{KLR} with
the Planck 1yr data.
In Sec.~\ref{obcon} we shall place observational bounds on
the same model as well as more general models from the
latest Planck temperature data combined with other data.

\section{Auxiliary vector modified $f(R)$ theories and
inflationary observables}
\label{mofR}

The discussion in Sec.~\ref{auximodel} can be extended to
more general auxiliary vector modified $f(R)$ theories.
In this section we shall perform such an analysis and then
provide the formulas of the primordial power spectra
of scalar and tensor perturbations generated during inflation.

\subsection{Auxiliary vector modified $f(R)$ theories}
\label{auxfRsec}

The auxiliary vector modification to $f(R)$ theories requires replacing
$R$ with $R + A_\m A^\m + \b \nabla_\m A^\m$.
Thus the model is given by the action (\ref{Sauxi}) with
\be
\cL = f(R + A_\m A^\m + \b \nabla_\m A^\m) \,.
\label{afr}
\ee
We rewrite this Lagrangian of the following form
\be
\cL = f(F) - \vf (F - R -  A_\m A^\m - \b \nabla_\m A^\m)\,.
\label{afr2}
\ee
Varying Eq.~(\ref{afr2}) with respect to $A^{\mu}$
and $F$, we obtain
\bea
A_\m &=& \frac1{2\vf} \b \nabla_\m \vf \,,
\label{Amre} \\
f_{,F}(F) &=& \vf\,.
\label{fFre}
\eea
Here and in the following, a comma in the lower index
denotes the partial derivatives with respect to scalar quantities
represented in the index, e.g., $f_{,F} \equiv \partial f/\partial F$.
The quantity $F$ depends on $\vf$ through the relation (\ref{fFre}).

Substituting Eq.~(\ref{Amre}) into Eq.~(\ref{afr2}),
it follows that
\be
\cL = \vf R - \frac{1}{4\vf} \b^2 \nabla_\m \vf \nabla^\m \vf
-\left[ \vf F(\vf)-f(F(\vf)) \right] \,.
\ee
This is equivalent to the BD theory with the same BD parameter
as Eq.~(\ref{omega}). Compared to Eq.~(\ref{Lpo}),
the scalar potential in the Jordan frame is generalized
to $V_J(\vf)=\vf F(\vf)-f(F(\vf))$.

Introducing the scalar field $\phi$ as Eq.~(\ref{phidef}) and
carrying out the conformal transformation
$\widetilde{g}_{\m\n}=\vf\,g_{\m\n}$,
we obtain the Einstein-frame
action (\ref{Ein}) with the scalar potential
\be
V(\phi)=\frac{M_{\rm pl}^2}{2} e^{-\sqrt{\frac{2}{3 \a}}\frac{\phi}{M_{\rm pl}}}
\left[ F - e^{-\sqrt{\frac{2}{3 \a}} \frac{\phi}{M_{\rm pl}}} f(F) \right]\,,
\label{poEin}
\ee
where $f_{,F}(F)=\vf=e^{\sqrt{\frac{2}{3 \a}} \frac{\phi}{M_{\rm pl}}}$.
The $\alpha$-attractor model, which corresponds to
the potential (\ref{alpo}), is a special case of a larger
class of auxiliary vector modified $f(R)$ theories.
When $\beta=0$ we have $A_{\mu}=0$ from Eq.~(\ref{Amre}), so
that the Lagrangian (\ref{afr}) recovers that of $f(R)$ theories.

\subsection{Inflationary observables}

Let us study inflation for the theories described by the action
(\ref{Sauxi}) with the Lagrangian (\ref{afr}).
We assume that the background is described by
the flat Friedmann-Lema\^{i}tre-Robertson-Walker
metric with the line element
$ds^2=-dt^2+a^2(t) \delta_{ij}dx^i dx^j$, where
the scale factor $a(t)$ depends on the cosmic time $t$.
We consider scalar and tensor metric perturbations
on this background.

In the Einstein frame the action is given by Eq.~(\ref{Ein})
with the potential (\ref{poEin}).
In Refs.~\cite{Fakir} it was shown that the inflationary observables
associated with linear scalar and tensor perturbations are
invariant under the conformal transformation.

The spectral index of scalar perturbations with the
primordial power spectrum ${\cal P}_{s}$ is defined by
$n_s \equiv 1+d\ln {\cal P}_{s}/d\ln k$, where $k$ is
a comoving wavenumber. We also introduce the
tensor-to-scalar ratio, as $r \equiv  {\cal P}_{h}/ {\cal P}_{s}$,
where $ {\cal P}_{h}$ is the primordial power spectrum of tensor
perturbations. Under the slow-roll approximation during
inflation, these observables are given by \cite{infreview}
\bea
{\cal P}_s &=& \frac{V^3}{12\pi^2 M_{\rm pl}^6 V_{,\phi}^2}\,,
\label{Ps}\\
n_s &=& 1-6\epsilon+2\eta\,,\label{ns}\\
r &=& 16 \epsilon\,, \label{r}
\eea
where
\be
\epsilon \equiv \frac{M_{\rm pl}^2}{2}
\left( \frac{V_{,\phi}}{V} \right)^2\,,\qquad
\eta \equiv \frac{M_{\rm pl}^2V_{,\phi \phi}}{V}\,.
\ee
As long as the slow-roll condition is satisfied, the analytic estimations 
(\ref{Ps})-(\ref{r}) are accurate enough to confront inflationary models 
with the CMB observations.
Defining the tensor spectral index as
$n_t \equiv d\ln {\cal P}_{h}/d\ln k$, the following
consistency relation holds \cite{infreview}
\be
r=-8n_t\,.
\ee

We define the number of e-foldings $N=\ln [a(t_f)/a(t)]$ in
the Jordan frame, where $a(t)$ and $a (t_f)$ are the
scale factors at the moments $t$ and $t_f$ respectively.
The lower index ``$f$'' represents the values at the end of
inflation. We identify the field value $\phi_f$
by the condition $\epsilon (\phi_f)=1$.

The number of e-foldings is a frame-independent quantity
by properly choosing the observer's reference frame \cite{Catena}.
On using the relations $\widetilde{a}=\Omega a$ and
$d \tilde{t}=\Omega dt$ \cite{conformal,Maeda}, the Hubble parameters
$\widetilde{H}=(d\widetilde{a}/d\tilde{t})/\widetilde{a}$ and
$H=(da/dt)/a$ in the two frames are related with each other as
$\widetilde{H}=[H+(d\Omega/dt)/\Omega]/\Omega$.
Since we are considering the choice (\ref{Omechoice}),
the number of e-foldings $N=\int_t^{t_f}Hdt$ can be
expressed as $N=\int_{\tilde{t}}^{\tilde{t}_f}\widetilde{H}d\tilde{t}
+\ln (\vf/\vf_f)^{1/2}$.
On using the slow-roll approximations
$3M_{\rm pl}^2 \widetilde{H}^2 \simeq V$ and
$3\widetilde{H} d\phi/d\tilde{t} \simeq -V_{,\phi}$
in the Einstein frame, it follows that
\be
N=\int_{\phi_f}^{\phi} \frac{V}{M_{\rm pl}^2 V_{,\phi}}d\phi
+\frac{1}{\sqrt{6\alpha}} \frac{\phi-\phi_f}{M_{\rm pl}}\,.
\label{efold}
\ee
The number of e-foldings associated with the CMB
temperature anisotropies corresponds to
$50 \lesssim N \lesssim 60$ \cite{infreview}.
On using Eq.~(\ref{efold}), the inflationary observables (\ref{ns})
and (\ref{r}) can be known as functions of $N$.

\section{Observational constraints from the latest CMB data}
\label{obcon}

We put observational constraints on several different inflationary
models that belong to the class of auxiliary vector
modified $f(R)$ theories.

We employ the bounds in the $(n_s, r)$ plane derived by the latest
Planck CMB temperature data (Temperature-Temperature (TT),
Temperature-E-mode (TE), E-mode-E-mode (EE) correlation
power spectra) and a first release of the B-mode
polarization data \cite{Planckcosmo}.
The Planck mission also performed the joint analysis
by taking into account the B-mode maps from BICEP2 and Keck Array
with the Planck maps  (BKP) at higher frequencies where the
emission of dust dominates \cite{BKP}. This study showed that
there is no statistical significant evidence for the detection of
primordial gravitational waves. Still, the latest BKP analysis put
tighter upper bounds on the tensor-to-scalar ratio $r$ than
those derived by the Planck data alone \cite{Planckinf}.

The likelihood analysis of the Planck mission is based on
expansions of the scalar and power spectra of the forms
${\cal P}_{s}(k)=A_{s}(k/k_*)^{n_s-1+(\alpha_s/2)\ln (k/k_*)+\cdots}$
and ${\cal P}_{h}(k)=A_{t}(k/k_*)^{n_t+(\alpha_t/2)\ln (k/k_*)+\cdots}$,
respectively, where $\alpha_{s,t}=dn_{s,t}/d\ln k$ are the runnings
of the spectral indices and $k_*$ is the pivot wavenumber.
Since there is no significant evidence for the large
deviation of $\alpha_{s,t}$
from 0, the standard slow-roll prediction of inflationary observables
is consistent with the CMB data.

\begin{figure}
\centering \noindent
\includegraphics[width=3.6in,height=3.5in]{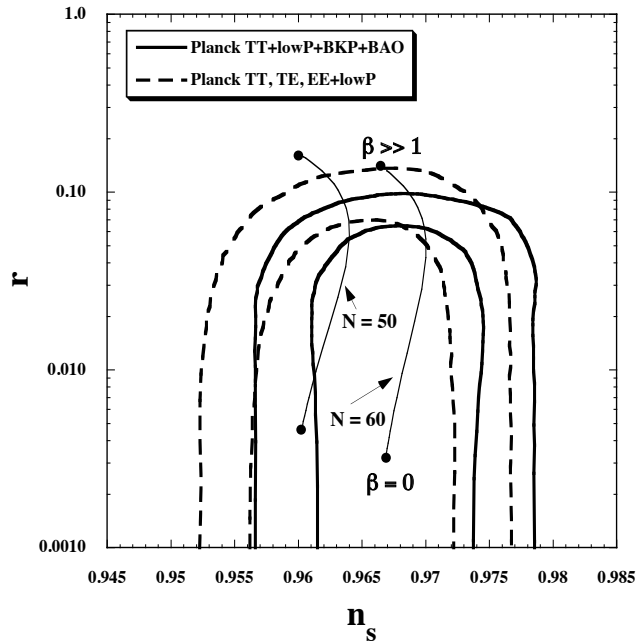}
\caption{The 68 \%\,CL (inside) and 95 \%\,CL (outside)
observational contours in the $(n_s, r)$ plane derived
from the joint data analysis of Planck TT+lowP+BKP+BAO
(thick solid) and Planck TT, TE, EE+lowP (thick dashed).
The pivot scale is chosen to be
$k_*=0.002$ Mpc$^{-1}$.
We also show the theoretical curves of the auxiliary vector
modified Starobinsky model (\ref{aS1}) as functions of $\beta$
for $N=50$ and $60$ (thin solid).}
 \label{fig2}
\end{figure}

In Fig.~\ref{fig2} we show the 68\,\%\,CL and 95\,\%\,CL
observational contours in the $(n_s, r)$ plane
derived by the latest Planck temperature
data as well as the BKP
and Baryon Acoustic Oscillations (BAO) data.
{}From the Planck TT, TE, EE and low-multipole temperature
polarization data (denoted as ``lowP''), the tensor-to-scalar
ratio $r$ is constrained to be $r<0.15$ (95\,\%\,CL) \cite{Planckinf}.
Combination of the BKP cross-correlation with the
Planck TT+lowP data gives a tighter
bound $r<0.08$ (95 \%\,CL).
Inclusion of the BAO data leads to the shift of $n_s$
toward larger values (as in the figure 1 of Ref.~\cite{Kuro}).
In the following we shall place observational constraints
on concrete auxiliary vector modified $f(R)$ models.

\subsection{Auxiliary vector modified Starobinsky model}
\label{ausec}

Let us begin with the model (\ref{aS1}), i.e.,
\be
f(F)=F+\frac{F^2}{6M^2}\,,
\ee
in the Lagrangian (\ref{afr2}). Since the potential in the
Einstein frame is given by Eq.~(\ref{alpo}),
the observables (\ref{Ps})-(\ref{r}) reduce to
\bea
{\cal P}_s &=& \frac{3\alpha}{128\pi^2} \left( \frac{M}
{M_{\rm pl}} \right)^2 \frac{(1-x)^4}{x^2}\,,\\
n_s &=&
1-\frac{8x(x+1)}{3\alpha(1-x)^2}\,,\label{ns2}\\
r &=&
\frac{64x^2}{3\alpha(1-x)^2}\,,\label{r2}
\eea
where $x \equiv e^{-\sqrt{2/(3\alpha)}\phi/M_{\rm pl}}$.
The number of e-foldings (\ref{efold}) reads
\be
N=\frac{3}{4}\alpha \left( \frac{1}{x}-\frac{1}{x_f} \right)
+\left( \frac{3}{4}\alpha-\frac12 \right) \ln \left(
\frac{x}{x_f} \right)\,,
\label{Nes}
\ee
where $x_f=(2\sqrt{3\alpha}-3\alpha)/(4-3\alpha)$.

When the parameter $\alpha=1+\beta^2/6$ is of the order of 1,
the inflationary epoch corresponds to the regime in which
$\phi$ is larger than $M_{\rm pl}$, i.e., $x \ll 1$.
Since $N \simeq 3\alpha/(4x)$ in this case, it follows that
\be
{\cal P}_s \simeq \frac{N^2M^2}{24\pi^2 \alpha M_{\rm pl}^2}\,,\qquad
n_s \simeq 1-\frac{2}{N}\,,\qquad
r \simeq \frac{12\alpha}{N^2}\,.
\ee
{}From the Planck normalization
${\cal P}_{s} \simeq 2.2 \times 10^{-9}$ \cite{Planckcosmo},
the mass scale $M$ is constrained to be
\be
\frac{M}{M_{\rm pl}} \simeq
1.3 \times 10^{-5} \sqrt{\alpha}
\left(\frac{N}{55}\right)^{-1} \qquad
{\rm for}~~\alpha=O(1)\,.
\label{masscon1}
\ee
In the presence of the coupling $\beta$,
both $M$ and $r$ are larger than those in the
the Starobinsky $f(R)$ model ($\alpha=1$).

In the limit that $\alpha \gg 1$, inflation occurs
in the region around $x=1$,
so the potential (\ref{alpo}) reduces to
$V(\phi) \simeq M^2 \phi^2/(2\alpha)$.
This means that, for increasing $\beta$, the observables
(\ref{ns2}) and (\ref{r2}) approach the values of
the quadratic potential, i.e.,
$n_s \simeq 1-2/N$ and $r \simeq 8/N$.
Since the scalar power spectrum is given by
${\cal P}_s \simeq N^2M^2/(6\pi^2 \alpha M_{\rm pl}^2)$,
the Planck normalization gives
\be
\frac{M}{M_{\rm pl}} \simeq
6.6 \times 10^{-6} \sqrt{\alpha}
\left(\frac{N}{55}\right)^{-1} \qquad
{\rm for}~~\alpha \gg 1\,.
\label{mcon2}
\ee
In this regime the mass scale $M$ is higher than that
for $\alpha=O(1)$.

In Fig.~\ref{fig2} we plot the theoretical curves
in the $(n_s, r)$ plane as functions of
$\beta$ (ranging $0 \le \beta \le 10^6$)
for $N=50$ and $60$.
The quadratic potential is outside the 95 \%\,CL
observational contours.
For $N=60$ the joint data analysis of
Planck TT+lowP+BKP+BAO gives
the following bounds
\bea
& &\beta<25 \qquad (68\,\%\,{\rm CL})\,,\\
& &\beta<66 \qquad (95\,\%\,{\rm CL})\,.
\eea
For $N=50$ the Starobinsky model ($\beta=0$)
is outside the 68 \%\,CL region
(mainly due to inclusion of the BAO data),
but it is still inside the 68\%\,CL contour constrained
by Planck TT, TE, EE+lowP.

\subsection{Power-law model}

We proceed to the power-law model given by
\be
f(F)=m^{2(1-n)} F^n\,,
\label{powerlaw}
\ee
where $m$ and $n$ are positive constants.
Since $F^{n-1}=\varphi/(nm^{2(1-n)})$ from Eq.~(\ref{fFre}),
the Einstein-frame potential (\ref{poEin}) reduces to
\be
V(\phi)=\frac{n-1}{2n^{n/(n-1)}}M_{\rm pl}^2m^2
\exp \left( -\frac{n-2}{n-1} \sqrt{\frac{2}{3\alpha}}
\frac{\phi}{M_{\rm pl}} \right)\,.
\ee
The positivity of the potential requires the condition $n>1$.
The power-law inflation ($a \propto t^p$ with $p>1$) \cite{powerlaw}
can be realized for $3\alpha>[(n-2)/(n-1)]^2$.
In this case we have
\bea
n_s &=& 1-\frac{2(n-2)^2}{3\alpha (n-1)^2}\,,\\
r &=& 8(1-n_s)\,.
\label{line}
\eea
The Harrison-Zeldovich spectrum corresponds to the
limit $n \to 2$ or $\alpha \to \infty$.

The theoretical values of $n_s$ and $r$ are
on the line (\ref{line}) in the $(n_s,r)$ plane.
This line is outside the 95 \%\,CL regions shown
in Fig.~\ref{fig2}, so the power-law model (\ref{powerlaw}) is
disfavored from the data.

\subsection{Generalization of the auxiliary vector
modified Starobinsky model}

Finally we study the following model
\be
f(F)=F+m^{2(1-n)} F^n\,,
\label{fmo}
\ee
where $m$ and $n~(\neq 2)$ are positive constants.
In this case the potential in the Einstein frame is given by
\be
V(\phi)=\frac{n-1}{2n^{n/(n-1)}}M_{\rm pl}^2 m^2
e^{-2\sqrt{\frac{2}{3\alpha}}\frac{\phi}{M_{\rm pl}}}
\left( e^{\sqrt{\frac{2}{3\alpha}}
\frac{\phi}{M_{\rm pl}}} -1 \right)^{\frac{n}{n-1}}\,.
\label{genepo}
\ee
Since we consider inflation in the regime $\phi>0$,
the positivity of the potential requires that $n>1$.

For given $n$ and $\beta$, we numerically
compute the field value $\phi_f$
at the end of inflation according to the condition $\epsilon(\phi_f)=1$.
From Eq.~(\ref{efold}) we identify the field value $\phi$
corresponding to $N=60$ and then evaluate the
observables (\ref{ns}) and (\ref{r}).

\begin{figure}
\centering \noindent
\includegraphics[width=3.6in,height=3.5in]{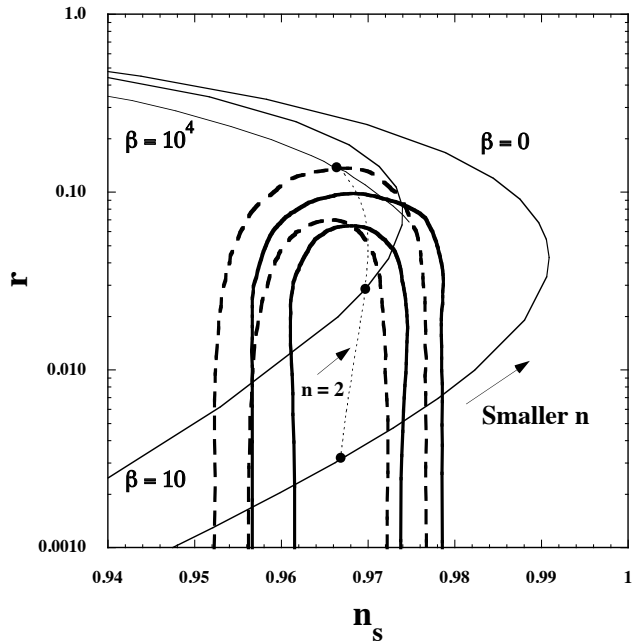}
\caption{
The theoretical curves of the model (\ref{fmo}) with
$N=60$ in the
$(n_s, r)$ plane for $\beta=0, 10, 10^4$.
These curves are plotted as as functions
of the power $n~(>1)$.
The theoretical prediction for the $n=2$
case is shown as a thin dotted curve.
For smaller $n$, the tensor-to-scalar ratio gets larger.
The 68\,\%\,CL and 95\,\%\,CL
observational contours are the same as those
plotted in Fig.~\ref{fig2}.
}
\label{fig3}
\end{figure}

Let us first discuss the case $\beta=0$.
Since inflation occurs in the regime where $\phi$ is bigger
than $M_{\rm pl}$, the large deviation of the power
$n$ from 2 spoils the flatness of the potential.
In Fig.~\ref{fig3} we plot the theoretical curve in the
$(n_s,r)$ plane for $N=60$ as a function of $n$.
The Starobinsky model ($n=2$)
is shown as a black circle.
For smaller $n$ the tensor-to-scalar ratio gets larger,
whereas the scalar spectral index reaches a maximum
value $n_s=0.991$ around $n=1.92$ and then turns
into decrease. When $n>2$, the inflationary observables
are particularly sensitive to the deviation from $n=2$
because of the appearance of the potential maximum
in the Einstein frame.
{}From the Planck TT+lowP+BKP+BAO joint analysis
we obtain the following bound
\be
1.980<n<2.015   \qquad (\beta=0)\,,
\label{bound1}
\ee
at 95\,\%\,CL.
Hence only the tiny deviation from $n=2$ is
allowed for the consistency with
the CMB data\footnote{This is also related to the fact that the
corrections like $\lambda_n R^n$ ($n>2$) to the Starobinsky model
$f(R)=R+R^2/(6M^2)$ need to be strongly suppressed
during inflation \cite{Huang}.}.
The mass scale $m$ constrained by the Planck
normalization is similar to that given in
Eq.~(\ref{masscon1}), with the correspondence
$m=\sqrt{6}M$ and $\alpha=1$.

In Fig.~\ref{fig3} we also plot the theoretical curve
for $\beta=10$ as a function of $n$.
The qualitative behavior of $n_s$ and $r$ with respect
to the change of $n$ is similar to
that for $\beta=0$.
The Planck TT+lowP+BKP+BAO joint analysis
gives the bound
\be
1.75<n<2.39   \qquad (\beta=10)\,,
\label{bound2}
\ee
at 95\,\%\,CL.
The wider range of $n$ is allowed
relative to the case $\beta=0$.
This reflects the fact that, for larger $\beta$, inflation
can occur in the regime where the quantity
$x=e^{-\sqrt{2/(3\alpha)}\phi/M_{\rm pl}}$
is not very much smaller than 1.
When $\beta=10$, for example, the order of $x$
satisfying the bound (\ref{bound2}) is typically 0.1
at $N=60$, whereas, for $\beta=0$,
$x$ is of the order of $0.01$
for $n$ ranging in Eq.~(\ref{bound1}).

In the limit $\beta \to \infty$ the epoch of inflation
corresponds to the regime $x \simeq 1$ and hence the
potential (\ref{genepo}) can be approximated as
\be
V(\phi) \simeq \frac{n-1}{2n^p} \left( \frac{2}{3\alpha}
\right)^{p/2} M_{\rm pl}^2 m^2 \left( \frac{\phi}{M_{\rm pl}}
\right)^p\,,
\ee
where
\be
p \equiv \frac{n}{n-1}\,.
\ee
This is equivalent to chaotic inflation with the power-law
potential $V(\phi) \propto \phi^p$, so
the inflationary observables are estimated as
\bea
{\cal P}_s &\simeq& \frac{c_n}{12\pi^2 p^2}
\left( \frac{m}{M_{\rm pl}} \right)^2
\left( \frac{\phi}{M_{\rm pl}} \right)^{p+2}\,,\\
n_s &\simeq& 1-\frac{p+2}{2N}\,,\label{powerlawns}\\
r &\simeq& \frac{4p}{N}\,,
\label{powerlawr}
\eea
where $c_n \equiv (n-1)[2/(3\alpha)]^{p/2}/(2n^p)$.
In the limits $n \to 1$ and $n \to \infty$ we have
$p \to \infty$ and $p \to 1$, respectively.
On using the relation
$\phi/M_{\rm pl} \simeq (2pN)^{1/2}$,
the Planck normalization constrains
the mass scale $m$, as
\be
\frac{m}{M_{\rm pl}} \simeq 5.1 \times 10^{-4}\,p\,
c_n^{-1/2}\,(2pN)^{-(p+2)/4}\,.
\ee

In Fig.~\ref{fig3} we plot the theoretical curve
for $\beta=10^4$ in the range $1.2 \le n \le 100$.
The values of $n_s$ and $r$ are very close to those
estimated from Eqs.~(\ref{powerlawns}) and
(\ref{powerlawr}).
Provided that $n>3.5$, the models are inside
the 95\,\%\,CL region constrained by the
Planck TT+lowP+BKP+BAO data.
However, even the linear potential $V(\phi) \propto \phi$
(i.e., $n \to \infty$) is marginally inside the 95\,\%\,CL contour,
so the models with $\beta \gg 1$ are not generally
favored from the CMB data.

\section{Conclusions}
\label{consec}

In this paper we showed that the auxiliary vector
modification to the Starobinsky model derived
by replacing $R$ with $R+A_{\mu}A^{\mu}+\beta \nabla_{\mu}A^{\mu}$
gives rise to the universal $\alpha$-attractor model proposed
in the context of supergravity.
Applying the same prescription to general $f(R)$ theories,
the resulting action is equivalent to that of BD theories
with the BD parameter $\omega_{\rm BD}=\beta^2/4$.
Under the conformal transformation to the Einstein frame, it is clear
that one scalar degree of freedom (a canonical field $\phi$)
propagates along the scalar potential.

For the potential with a sufficiently flat region the scalar
degree of freedom $\phi$ not only leads to inflation at the background level,
but also the field perturbation $\delta \phi$ can be
the source for primordial density perturbations relevant to the CMB temperature
anisotropies. Using the invariance of scalar/tensor perturbations
under the conformal transformation, the inflationary observables
in auxiliary vector modified $f(R)$ theories are simply given by
Eqs.~(\ref{Ps})-(\ref{r}) with the number of e-foldings (\ref{efold}).

In light of the recent release of the Planck temperature and
polarization data, we placed observational constraints on inflationary
models in the framework of auxiliary vector modified $f(R)$ theories.
We studied three different models:
(i) $f(F)=F+F^2/(6M^2)$, (ii) $f(F)=m^{2(1-n)}F^n$, and
(iii) $f(F)=F+m^{2(1-n)}F^n$ ($n \neq 2$), where
$F=R+A_{\mu}A^{\mu}+\beta \nabla_{\mu}A^{\mu}$.

The model (i) is equivalent to the $\alpha$-attractor model
with the correspondence $\alpha=1+\beta^2/6$,
which recovers the Starobinsky model for $\beta=0$.
{}From the joint data analysis of Planck TT+lowP+BKP+BAO
the parameter $\beta$ is constrained to be
$\beta<25$ (68\,\%\,CL) for $N=60$ (see Fig.~\ref{fig2}).
The model (ii) gives rise to the exponential potential
in the Einstein frame, in which case the theoretical line
in the $(n_s,r)$ plane is outside the 95\,\%\,CL
observational contours.

For the model (iii) with $\beta=0$, the power $n$ is
constrained to be in the narrow range around $n=2$, i.e.,
$1.980<n<2.015$ (95\,\%\,CL) from
the Planck TT+lowP+BKP+BAO data.
With increasing $\beta$, the allowed region of $n$
tends to be wider because inflation occurs for
$x=e^{-\sqrt{2/(3\alpha)}\phi/M_{\rm pl}}$ not very
much smaller than 1. In the limit $\beta \to \infty$
the theoretical values of $n_s$ and $r$ are the same
as those in chaotic inflation with the potential
$V(\phi) \propto \phi^{n/(n-1)}$, in which case
the model is marginally inside the 95\,\%\,CL
observational contour for $n>3.5$.

The issue of super-symmetrization of the auxiliary vector 
modified Starobinsky model would be interesting. 
However this is not an easy task, so we leave it for future investigation. 
{}From the observational side, the possible detection of primordial 
gravitational waves will be able to clarify whether or not the Starobinsky
model and the auxiliary vector modified $f(R)$ models
are observationally favored.
We hope that we can approach the origin of inflation in the
foreseeable future.

\section*{Acknowledgements}

MO would like to thank to Eric Bergshoeff, Renata Kallosh,
Andrei Linde, and Diederik Roest for useful discussions.
The work of ST is supported by the Grant-in-Aid for Scientific
Research from JSPS (No.~24540286)
and by the cooperation programs of Tokyo University of Science
and CSIC. The work of YP was supported
in part by DOE grant DE-FG02-13ER42020.


\appendix
\section{The onset of radiation-dominated era}
\label{appendix} 

We estimate the time $t_r$ at which the energy density of 
radiation dominates over that of the field $\phi$ for 
the auxiliary modified Starobinsky model. 
In the Starobinsky model ($\alpha=1$), 
this issue was already addressed in Refs.~\cite{reheating2,fRreview}. 
During the oscillating stage of inflaton the potential 
(\ref{alpo}) is approximately given by $V(\phi) \simeq m^2 \phi^2/2$, 
where $m \equiv M/\sqrt{\alpha}$. Hence the discussion for the model 
$\alpha \neq 1$ is analogous to that given in Refs.~\cite{reheating2,fRreview} 
after the replacement of $M$ with $m$.
In what follows we estimate the time $t_r$ briefly.

To study the particle production during reheating, let us consider 
a massless canonical scalar field $\chi$ in the Jordan frame.
We express the quantum field $\chi$ in terms of the 
Heisenberg representation:
\be
\chi (t, {\bm x})=\frac{1}{(2\pi)^{3/2}} 
\int d^3k \left( \hat{a}_k \chi_k (t) e^{-i {\bm k}\cdot {\bm x}}
+\hat{a}^{\dagger}_k \chi_k^* (t) e^{i {\bm k}\cdot {\bm x}}
\right)\,,
\ee
where $\hat{a}_k$ and $\hat{a}^{\dagger}_k$ are annihilation 
and creation operators, respectively. 
The rescaled field $u_k=a \chi_k$ obeys the equation of motion
\begin{equation}
\frac{\mathrm{d}^2 u_k}{\mathrm{d}\eta^2}+
k^2 u_k=U(\eta)u_k\,,
\label{ukeq}
\end{equation}
where $U (\eta)=a^2 R/6$, and 
$\eta=\int a^{-1} \mathrm{d}t$ is the conformal time.
The time-dependent term on the r.h.s. of Eq.~(\ref{ukeq})
leads to the production of $\chi$ particles with the initial 
vacuum state described by the solution 
$u_k^{(i)}=e^{-ik \eta}/\sqrt{2k}$.

The energy density $\rho_{r}$ of the field $\chi$ is associated 
with the Bogoliubov coefficient $\beta_k=
-\frac{i}{2k} \int_{0}^{\infty} U(\bar{\eta})e^{-2ik\bar{\eta}}d\bar{\eta}$, 
as $\rho_{r}=\frac{g_*}{(2\pi)^3a^4} \int_0^{\infty} 
4\pi k^2 dk \cdot k |\beta_k|^2$, where $g_*=O(100)$ is 
the number of relativistic degrees of freedom.
During reheating the Ricci scalar evolves as 
$R=O(1) \frac{m}{t-t_{\rm os}} \sin [m(t-t_{\rm os})]$ in the 
regime $m(t-t_{\rm os}) \gg 1$, 
where $t_{\rm os}$ is the time at the onset of reheating. 
Taking the time average of oscillations of $R$, the energy density 
of created particles can be estimated as \cite{reheating2,fRreview} 
\be
\rho_r ={\cal C} \frac{m^3}{t-t_{\rm os}}\,,
\label{rad}
\ee
where ${\cal C}$ is a coefficient of the order of 1. 
The scale factor evolves as $a \propto (t-t_{\rm os})^{2/3}$ during the 
oscillating phase of $\phi$, so the evolution of the Hubble parameter squared 
is given by 
\be
H^2=\frac{4}{9(t-t_{\rm os})^2}\,.
\label{Hden}
\ee

The radiation density (\ref{rad}) decreases slowly relative to $H^2$ (which 
is proportional to the field density $\rho_{\phi}$). 
The onset of radiation-dominated epoch ($t=t_r$) is identified by 
the condition $3M_{\rm pl}^2 H^2=\rho_r$, i.e., 
\be
t_r-t_{\rm os}=\frac{4}{3{\cal C}} 
\frac{M_{\rm pl}^2}{m^3}=\frac{4}{3{\cal C}}
\alpha^{3/2}\frac{M_{\rm pl}^2}{M^3} \,.
\ee
Using the observational constraint (\ref{mcon2}), 
which is valid in the regime $\alpha \gg 1$, we obtain
\be
t_r-t_{\rm os} \simeq \frac{5 \times 10^{15}}{\cal C}
\left( \frac{N}{55} \right)^3 \frac{1}{M_{\rm pl}}\,.
\ee
Substituting the values ${\cal C}=O(1)$ and $N=55$ it follows that 
$t_r-t_{\rm os} \approx 10^{-27}$~sec. 
For $t>t_{r}$ the inflaton energy density $\rho_{\phi}$ becomes negligible 
relative to $\rho_{r}$, so it does not affect the thermal history of the Universe 
after the onset of the radiation era.

\end{document}